# Co-mapping Cellular Content and Extracellular Matrix with Hemodynamics in Intact Arterial Tissues Using Scanning Immunofluorescent Multiphoton Microscopy

Yasutaka Tobe, Anne M. Robertson, Mehdi Ramezanpour, Juan R. Cebral, Simon c. Watkins, Fady T. Charbel, Sepideh Amin-Hanjani, Alexander K Yu, Boyle C. Cheng, Henry H. Woo

**Abstract:**

Deviation of blood flow from an optimal range is known to be associated with the initiation and progression of vascular pathologies. Important open questions remain about how the abnormal flow drives specific wall changes in pathologies such as cerebral aneurysms where the flow is highly heterogeneous and complex. This knowledge gap precludes the clinical use of readily available flow data to predict outcomes and improve treatment of these diseases. As both flow and the pathological wall changes are spatially heterogeneous, a crucial requirement for progress in this area is a methodology for co-mapping local data from vascular wall biology with local hemodynamic data. In this study, we developed an imaging pipeline to address this pressing need. A protocol that employs scanning multiphoton microscopy was designed to obtain 3D data sets for smooth muscle actin, collagen and elastin in intact vascular specimens. A cluster analysis was developed to objectively categorize the smooth muscle cells (SMC) across the vascular specimen based on SMC density. In the final step in this pipeline, the location specific categorization of SMC, along with wall thickness was co-mapped with patient specific hemodynamic results, enabling direct quantitative comparison of local flow and wall biology in 3D intact specimens.

**Introduction**

It has been well established that deviations in the quantitative and qualitative features of blood flow from an optimal range are associated with the development and progression of vascular diseases such as atherosclerosis and cerebral aneurysms.[1-4] However, these diseases involve complex changes to multiple wall components and fundamental open questions remain regarding how flow features are directly coupled to even commonly seen pathological changes such as inflammation, apoptosis of smooth muscle cells, fibrosis, and formation of atherosclerotic plaques. While the influence of specific aspects of flow such as magnitude of wall shear stress on vascular biology have received attention in relatively simple rectilinear flow fields, many pathological flows, such as those found in cerebral aneurysms, are vastly more complex that these idealized flows.[5,6] As a result, for many diseases, controversy remains over which aspect of the abnormal flow is most important in driving pathological changes.[6-8]

If the relationship between flow and pathological wall changes were better understood, it would be possible to design more effective treatments. For example, flow metrics for patients with brain aneurysms are readily available from image-based flow studies and are believed to have great potential to improve assessment of rupture risk and to guide endovascular therapies.[5,7] However, the lack of knowledge as to how the abnormal flow in the aneurysm is related to degenerative wall changes precludes the clinical use of this data.[7] The design and use of vascular implants is another area where there is a pressing need to better understand the coupling between altered flow and changes to the arterial wall. For example, computational fluid dynamic studies of the flow experienced by saphenous veins after coronary artery bypass grafting (CABG) have identified associations between flow metrics and graft failure. A mechanistic understanding of this association would provide

direction to redesign grafts and improve the low patency rates of 55-60% at ten year follow up.[9,10]

A central impediment to progress in this area is the lack of methodologies for directly relating local changes in vascular wall pathologies to local flow parameters.[7,11] Some vascular pathologies, such as lipid pools and calcification, become large enough to be detected in vivo using non-invasive modalities such as computational tomography (CT), magnetic resonance imaging (MRI) and ultrasound. Such data can then be directly related to flow parameters derived from computational fluid dynamic simulations in 3D image-based reconstructions of the patient vasculature. However, these imaging modalities cannot currently provide information about cellular content in the wall, nor the state of the extracellular matrix (e.g. collagen and elastin fibers). Knowledge of the state of ECM is important as a determining factor for the mechanical behavior of the wall and degeneration of ECM can lead to soft tissue failure. Furthermore, in early stages of disease the size of lipid pools and calcification are too small to be detected by standard clinical scanners. Rather, data on these wall changes must necessarily be obtained ex-vivo from harvested tissue samples.[12]

Table 1 compares the attributes of ex vivo imaging methodologies commonly used for soft tissue samples. An important consideration in selecting the modality is requirements for sample preparation. Optical microscopy, the most common method for imaging cells and extracellular matrix in soft tissues, is performed on fixed tissue specimens that are sectioned and stained using classical histological methods or immunohistochemistry. The discontinuous nature of these data sets limits the possibility of analyzing the 3D network structure the ECM, such as collagen and elastin, important feature for biomechanical studies.[13-16] Recently, Niemann et al. developed an approach to create 3D histological models from the stacks of 2D sections.[17,18] This is an important step forward for enabling the implementation of wall biology into biomechanical studies, though the accuracy of the 3D model is still limited by the destructive nature of the 2D serial sectioning process and the discontinuity of the data between slices.

Table 1: Attributes of methodologies for ex vivo imaging of vascular

|  | Optical Microscopy | Multiphoton Microscopy (MPM) | Confocal Microscopy (CFM) | Micro-CT |
|---|---|---|---|---|
| Does not require fixation |  | ✓ | ✓ | ✓ |
| Cell distribution visible | ✓ | ✓ | ✓ |  |
| Co-staining possible | ✓ | ✓ | ✓ |  |
| Can image intact specimens (without sectioning) |  | ✓ | ✓ | ✓ |
| Deep imaging (>1 mm) |  |  |  | ✓ |
| Moderate depth imaging (200-500 um) |  | ✓ |  | ✓ |
| Moderate to large Field of View (>500 um) | ✓ |  |  | ✓ |

In cases, where intact samples can be imaged, these limitations can be avoided. Confocal (CFM) and multiphoton (MPM) microscopy are both commonly used approaches for imaging intact soft tissues specimens in 3D without damaging or deforming the tissue. MPM has the advantage of deeper imaging depth of 300 um to 500 um in dense collagenous tissues in comparison to 200 um to 300 um for CFM.

Regardless of the microscopy approach, the challenge remains of relating the local flow dynamics to the histological data sets. In a study of the role of flow dynamics in the early stages of cerebral aneurysm formation, Tremmel et al. introduced a methodology to co-map virtual slices of the hemodynamic flow field at arterial bifurcations to 2D histological data sets in that region.[19] Using this approach, they were able to demonstrate that a combination of elevated wall shear stress and wall shear stress gradient led to degradation of the internal elastic lamina and medial thinning.[6,19,20] The recent work by Niemann et al. extended this methodology to 3D setting by creating 3D histological models formed from the stacks of 2D histological sections.[17] This approach enables 3D reconstructed data from high resolution optical microscopy to be co-mapped to 3D hemodynamic data sets, though the previously mentioned artifacts associated with this destructive approach remain.

In parallel with these studies, our group developed a pipeline to co-map flow parameters to multiphoton data in intact specimens including collagen fiber architecture, wall thickness, and calcification.[21,22] This pipeline did not include data on the cellular content in the wall and was limited by the small field of view from the multiphoton microscope (~ 500 µm x 500 µm). In this study, we build on this prior work to introduce a new imaging pipeline that overcomes the limitations in size of the imaging region as well as the need to image cellular distribution in the wall. In particular, this approach utilizes second harmonic generation and immunofluorescent co-staining to non-destructively co-image cellular content and ECM structure in large three dimensional sample volumes, made possible with scanning MPM. By using scanning immunufluroescent multiphoton microscopy (SI-MPM) it is possible to obtain the distribution of SMC and collagen organization across the entire intact sample. Since vasculature pathologies can be highly heterogeneous and intramural cells such as vascular smooth muscle cells have an important role, these advances have important scientific implications.

A common step in the analysis of the data generated from these studies is categorization of the state of the vascular pathology. For example, a seminal work on cerebral aneurysms categorized each aneurysm tissue based on the SMC density and organization within the tissue in order to rank the state of the aneurysm wall.[23] This approach was based on 2D histology and assigned a single ranked type to each specimen based on a sampling of 2D slices. The methodology introduced in the present work provides an opportunity to provide local assessment of the state of the wall *across* the entire intact specimen. We introduce an objective method (based on a cluster analysis) to analyze this distributed data and generate a map of wall type across each specimen, rather than a single wall type for each specimen. This local information can be compared with the local flow information.

Here, we focus on SMC, elastin fibers and collagen fibers within specimens from human brain arteries and aneurysms, though the approach can be applied to other soft biological tissues and cell types. SMCs can be influenced by fluid shear stress both indirectly, through the endothelial cell response to flow and as well as directly through interstitial flow.[24]. In healthy vessels, the SMCs are exposed to slow intersitial flow across the wall and largely protected from the faster flow along the vessel wall. However, the endothelium can be damaged or absent in pathological tissues such as cerebral aneurysms, atherosclerotic vessels or following endovascular treatment. Therefore, in these tissues, the SMC can be influenced more directly by the flowing blood. [24] [25] [26] [27]

# Results

## Scanning Immunofluorescent Multiphon Microscopy (SI-MPM) imaging captures autofluorescing and immunofluorescing signals more effectively than CFM

Capturing the 3D nature of the extracellular matrix (ECM) structure along with cellular contents in arterial tissues requires an appropriate whole mount staining protocol coupled with a 3D imaging modality with the capacity to image large areas with moderate or high imaging depth. To accomplish this, we applied immunofluorescent stainings to whole samples and imaged them with the scanning multiphoton microscope. Here, we identify SMC by staining actin fliaments within SMCs ($\alpha$SMA). In addition to the conventional Z moving stage, the scanning multiphoton microscope has an XY moving stage, making it possible to capture immunofluorescent and/or autofluorescent signals over a larger area, equivalent to that possible with a ribbon CFM, while maintaining the imaging depth of MPM.

First, we illustrated the capacity for stacking autofluorescing and immunofluorescing signals obtained from scanning MPM of arterial specimens without interference. A cerebral artery from cadaver was cryosectioned, immunostained, and scanned to assess the capability of MPM for co-imaging collagen, elastin and SMC in 2D sections (Fig. 1a). Using MPM it was possible to clearly image the autofluorescing signal from collagen fibers (Ch1), internal elastic lamina (Ch1, Ch2, white arrow) and SMC actin (Ch2, green) in a single scan. The adventitial collagen has a higher signal intensity than the medial collagen (Ch1). The stacked image confirms both collagen and SMC actin are uniquely fluorescing in each channel and can therefore be distinguished. The thick internal elastic lamina (IEL) was present in both channels (yellow in stacked image).

Second, we evaluated the ability of MPM to capture both autofluorescing and immunofluorescing signals in 3D with equivalent or higher quality than the conventional CFM. 3D scans using CFM and SI-MPM on neighboring cylindrical sections of a small perforating artery from a middle cerebral artery (MCA) were compared, (Fig. 1b, 3D). The field of view (FoV) of CFM was 500 μm x 500 μm and imaging depth was 120 μm. In contrast, for the SI-MPM, the entire surface of the sample could be scanned to cover an imaging area of FoV of 4340 μm x 1045 μm with an imaging depth of 914 μm, (Fig. 1b, 3D, SI-MPM). The difference in imaging depth is clear from comparison of the virtual slices of each of the 3D volumes (Fig. 1b, YZ virtual slice). In order to compare the images with similar area and magnification, a 500 μm x 500 μm subregion of the SI-MPM data was selected (Fig. 1b, 3D, SI-MPM, white dotted square). The Z projection images for both CFM and SI-MPM were also analyzed. As a reference, the negative control for αSMA showed no SMC actin under CFM for any channel (Fig. 1b, CFM, NC, all rows). Similarly, the negative control of SI-MPM did not show any SMC actin signal, though both the autofluorescing collagen fibers (Fig. 1b, SI-MPM, NC, all rows) and the undulations of the IEL (Fig. 1b, SI-MPM, NC, Ch 2) were visible (not possible under CFM). The autofluorescing collagen signal was still successfully captured by SI-MPM in the stained sample (Fig 1b, SI-MPM, SMC actin, Ch 1). While the SMC actin were visible under both CFM (Fig 1b, CFM, SMC actin, Ch 2) and SI-MPM (Fig 1b, SI-MPM, SMC actin, Ch 2), the SMC actin in CFM showed less clarity than under SI-MPM. In the SI-MPM images it was even possible to identify the circumferentially aligned SMC actin in the artery (Fig 1b, SI-MPM, SMC actin, Ch 2). Stacked images (Fig 1b, SI-MPM, SMC actin, stacked) show the SI-MPM successfully captured both the immunofluorescent signal of SMC actin (green) and autofluorescent signal for collagen (red) independently, as was possible for the 2D sample. In SI-MPM , by adjusting the threshold and contrast of Ch 2, it was possible to visualize SMC actin and IEL, individually or collectively. In particular, the signal from elastin was clearly visible when the thresholding and contrast were adjusted to include lower intensity signals, (Fig. 1c, dim green).

**SI-MPM can co-image SMC actin and collagen in high detail throughout the volume of thin, intact pathological arterial specimens**

The ECM of pathological tissues can differ from non-pathological arteries affecting, for example, the types and density of collagen fibers, which can in turn influence the quality of MPM imaging.[28] We therefore, evaluated the SI-MPM protocol in pathological arterial tissues. In particular, we imaged two cerebral aneurysm samples with different wall thicknesses. The first sample (Fig. 2a) had a maximum wall thickness of 270 μm and had an average wall thickness of 97 ± 56 μm (sample size ~2 x 3 mm) based on the 3D reconstruction of micro-CT scan of tissue sample. SMC actin (Fig. 2a, Ch2) was found over a majority of the sample, though small regions lacked SMC actin. A virtual slice in the orthogonal plane at location i demonstrated MPM was capable of imaging through entire thickness of this sample (Fig. 2a-i). Further, SI-MPM captured the heterogeneity of the collagen structure and density across 3D volume. In the Z projection (Fig. 2a, Ch1) the heterogeneity in signal intensity can be seen in the lower half of the image. In the virtual XZ slice, regions of sparse collagen can be seen at the center of the wall.

The average wall thickness of the second sample was 549 ± 279 μm and was 750 μm in the thickest region (Fig. 2b, Wall thickness). Though the entire projected area could be imaged, it was not possible to image through the entire thickness. Therefore, scans were performed from both luminal and abluminal sides. From the lumen side, the collagen appeared amorphous (Fig. 2b, Ch1) rather than as the distinct collagen fibers seen in the thin sample (Fig. 2a, Ch 1). SMC actin was largely present across the entire sample (Fig. 2b, Ch2 lumen) with only a few spots lacking signal. The region dependent absence of SMC actin was also seen in a virtual slice at location i (Fig. 2b-i). As previously noted, the sample thickness exceeded the imaging depth of the microscope and as a consequence the collagen signal was lost within the tissue (Fig. 2b, white arrow). In a virtual slice at location ii (Fig. 2b-ii), the SMC actin signal was seen to be concentrated just below the lumen. When imaged from the abluminal side, a complex network of thick collagen fibers was seen across the sample (Fig. 2b, Ch1, ablumen), whereas the SMC actin signal was largely absent (Fig. 2b, Ch2, ablumen). Virtual XZ slices at location iii and iv both showed loss of signal beyond 100 $\mu$m when imaged from the albumen side (Fig. 2b-iii, iv, white arrow).

**Use of SI-MPM images for computational pathology**

An important consideration in analysis of microscopy data is development of protocols that provide objective quantitative and categorical results, avoiding user bias and speeding analysis. Here, we developed an automated cluster analysis for the Z projection of SMC actin data in order to objectively categorize regions of different SMC actin type across pathological samples. In particular, three wall categories were introduced based on SMC actin density (Fig. 3a). Type 1, 2, and 3 regions had high, moderate and low density, respectively. The wall typing was illustrated in four aneurysm specimens (Fig. 3b) and the distribution of types across each sample was analyzed from this data (Fig. 3c). SMC actin type was heterogeneous within and between samples. For example, the aneurysm tissue for Case 1 was predominantly Type I, though it displayed all three wall types. The Type I region was separated into two subregions by an interface with low or moderate density SMC actin. The Case 2 aneurysm sample was dominated by regions with either moderate or low density SMC actin. In contrast, the aneurysm from Case 3 was endowed with rich levels of SMC actin across 93% of the sample with small scattered Type 3 areas. While the majority of the aneurysm sample for Case 4 lacked SMC actin signal (Type 3), the central region was well endowed with SMC actin

**Mapping SI-MPM data to vascular models for studies of the role of hemodynamics in wall pathology**

To determine the possible role of abnormal hemodynamics on pathologoical changes in the vascular wall, it is essential to co-map wall pathology data with hemodynamic data. Here we demonstrate a methodology to co-map the SI-MPM biological data from an aneurysm sample to flow parameters obtained from computational fluid dynamic simulations, performed using the corresponding patient specific vasculature. To co-map these two data sets, we created a 3D in silico (or virtual) model of the harvested tissue sample using micro-CT data. The SI-MPM data was first mapped to this in silico model which could then be mapped to the 3D vasculature. Using this approach the wall pathology data (SMC actin type, collagen fiber orientation) can be directly related to patient specific hemodynamic data (e.g. wall shear stress).

Alignment of the harvested specimen to its original position in the 3D vasculature is the most challenging aspect of this process (Fig 4a). Access to surgical video with images of the aneurysm before (Fig 4a, Surgical view) and after specimen harvest (Fig 4a, Tissue harvest) are necessary. Geometric landmarks on the physical sample that are apparent in the surgical video as well as landmarks applied to the tissue during surgery using surgical marking pens (Xodus Medical Inc.) (white arrow) are used to determine the orientation and origin of the harvested specimen on the 3D reconstructed aneurysm. These landmarks are also visible in the dissection scope image of the specimen. In the right panel of Fig. 4a, the 3D in silico (or virtual) model of the harvested tissue (orange) that was created from the 3D micro-CT data is shown aligned on the 3D reconstructed vascular model.

The wall pathology data obtained from SI-MPM (here, the 2D data set for SMC actin density type) is projected onto the lumen face of the in silico 3D specimen model (Fig. 4b, left). Wall thickness maps for the sample (Fig. 4b, middle) can directly be compared with this pathology data either qualitatively or directly based on subregions in the virtual sample. Both the SMC actin data and then wall thickness data can in turn be mapped onto the vasculature (Fig. 4b, right) using the prior alignment shown in Fig. 4a. In the right panel, the abluminal side of the in silico model is outward and made opaque so the colored wall regions are visible.

The mapped pathology data, can then be directly related to the patient specific hemodynamic results that were obtained in the same 3D vasculature, (Fig. 4c). In particular, flow results at systole for Case 4 are shown in Fig. 4c (row 1). In row 2, the SMC actin wall type data obtained from Case 4 (Fig. 3), is co-mapped with the flow data. Using this mapping, biological and flow data can be quantitatively compared pointwise across the surface. Even on gross observation, the aneurysm specimen can be seen to be subjected to a vortex flow with uniformly low wall shear stress magnitude (largely below 15 dyne/cm$^2$, even in systole). The vortex center impacts the tissue sample (Fig. 4c, row 2, Vortex streamline), and the vortex core line is located near the thickened bleb of the aneurysm (Fig. 4c, vortex core line) where the specimen had low SMC actin signal. The scattered regions of Type 1 and 2 were found away from the vortex (Fig. 4c, vortex streamline, white arrow).

**Discussion**

Complex patterns of blood flow within the cardiovascular system generate abnormal wall shear stress fields that are known to drive pathological changes to the vascular wall. [4,11,29] In some diseases, such as cerebral aneurysms, both the flow and the wall pathology can be highly heterogeneous,[5,7] necessitating a methodology to spatially register local cellular content in the wall to flow dynamics. In this work, we addressed this pressing need by designing a pipeline that employs scanning immunofluorescent multiphoton microscopy to non-destructively obtain comprehensive 3D images of both ECM and SMC across intact vascular specimens. This framework includes a methodology to objectively analyze this cellular data using a cluster analysis, generating categorical data on SMC density across the specimen. The location specific data on SMC density, along with wall thickness could then be mapped to patient specific hemodynamic results enabling direct quantitative comparison of focal flow parameters and wall biology in 3D intact specimens. In this work, we focused on cerebral aneurysm tissue, though the framework can be more generally applied to other vascular tissues.

An important contribution of the present study is the introduction of a method to objectively categorize the SMC in the vascular wall. We identified high, medium, and low-density subregions in the 2D projected SMC actin data sets using unsupervised machine-learning-based image segmentation algorithms. There are numerous supervised machine-learning-based segmentation algorithms ranging from less-sophisticated logistic regression classifiers to highly complex deep-learning-based convolutional neural networks. Supervised machine learning segmentation algorithms generally require large labeled datasets as the ground truth for training purposes and also have high computational costs associated with the training process. In contrast, the unsupervised machine learning segmentation algorithms, ranging from very simple image thresholding methods to density-based clustering algorithms such as DBScan, do not require labeled datasets to be trained and are substantially faster than the supervised algorithms. For this work, an unsupervised cluster analysis algorithm was found to be effective for categorizing SMC actin density. The application of machine learning to the field of pathology is an area of growing focus as it provides a fast and repeatable means for objectively analyzing and categorizing specimens, though open challenges remain for routine clinical implementation.[30] The clinical value of these algorithms has been proven in a number of applications, such as detection of metastases of lymph nodes, demonstrating the predictive capability can be higher than even clinical experts. [31]

While the cerebral aneurysm wall has long been recognized as heterogenous,[32] it is only more recently that the structural importance of the wall heterogeneity has been considered.[12,21,22,28,33-36] The relevance of the structural heterogeneity to the rupture process is of clinical importance for a variety of vascular diseases including atherosclerosis and aneurysms. Collagen, the central load bearing component in vascular tissues is maintained by the SMC and fibroblasts. [13,37] Therefore, the state of the SMC within the wall is of primary importance in determining wall strength. Frösen et al. identified four cerebral aneurysm wall types based on cell content and organization, ranked from A to D. Type A had linearly organized SMC, Type B disorganized SMC, Type C and D were hypocellular, with the D walls being thin. A significant association was found between wall type and rupture, with increasing likelihood of rupture from A to D. One limitation in this prior work is that only a single wall type was assigned to each sample, based on analysis of a subset of histological sections. This may explain why even 42% of Type A walls were ruptured. Using the framework introduced here, it is possible to assess the state of SMC across the specimen and thereby extend the aneurysm wall classification to a local level while also eliminating the need for subjective analysis. With regards to the Type A and B categories, here, we chose to assess SMC actin density without a prior assumptions about the relationship between density and SMC actin aligntment. For the small number of samples studied here, SMC actin in Type I SMC regions were found to be elongated in shape while SMC actin in Type II regions tended to be spherical, Fig.2(a). This relationship between SMC actin density and alignment will be investigated in a larger study. The Type III regions in the current study can be further subdivided into Type C and D of the Frösen typing using the wall thickness data from the same intact tissue specimens Fig. 4(b). For example, the Type III wall in Fig. 3 is all of Type C.

An important factor in the success of the current approach to co-map flow and wall components is the non-destructive comprehensive nature of the SI-MPM imaging. Important prior work introduced a pipeline to reconstruct 3D composite models of vascular wall from serial sectioned histological data sets using machine learning algorithms to differentiate wall regions in each 2D slice.[18] However, a challenge with these data sets is the inherent physical distortion of the sections along with the requirement for fixation. The discontinuous nature of the data means that some localized heterogeneities, including calcification particles and interfaces between wall regions may be lost. Furthermore, it is difficult, if not impossible, to explore the 3D nature of the collagen fiber organization in serial sectioned data sets. An advantage of this more destructive approach is that it does not suffer from limitations in imaging depth, a consideration when using SI-MPM on thicker specimens such as Sample 2 in Fig. 2. A variety of optical clearing methods have been developed[38] that increase the imaging depth by an order of magnitude by changing the refractive index of the sample. Such approaches can be used prior to SI-MPM analysis, though some protocol depenendent tradeoffs can exist due to the clearing process, such as swelling, removal of lipids, and dehydration of the sample.

Lastly, we demonstrated an approach for co-mapping CFD and SI-MPM data, allowing a direct comparison of regional hemodynamics and wall biology. There are few prior studies of the relationship between wall heterogeneity and flow in cerebral aneurysm pathology.[3,11,22,35] Of relevance is a study of the association between flow and inflammatory markers for 18 aneurysm samples. As tools were not available to directly investigate local associations, the authors chose to asses the association between heterogeneity in fluid wall shear stress (WSS) across the aneurysm dome and heterogeneity in the numbers of inflammatory cells in representative serial sections.[35] They authors conjectured the significant association they found arose from flow driving wall biology at a local level, "suggesting that interactions leading to the association of high WSS and inflammation might happen focally at the wall." An important accomplishment of the present work is the introduction of an approach for comprehensive imaging and quantitative assessment across the specimen, making it possible to directly evaluate conjectures of this kind. We illustrated this approach for a comparison of local SMC and hemodynamic characteristics. This same approach can be applied to other cells (e.g. endothelial cells, inflammatory cells) as well as other biomechanical field variables such as intramural stress parameters and even geometric parameters such as local wall curvature.

**Methods**

**Acquisition of vascular specimens**

A circle of Willis (CoW) was obtained at autopsy from a 65 years old male, (Brain Bank of the University of Pittsburgh). The right middle cerebral artery (MCA) with perforators was resected from the CoW and fixed within 24 hours in 4% paraformaldehyde. The CoW had no obvious atherosclerosis. Specimen harvest and handling followed a protocol approved by the Committee for Oversight of Research and Clinical Training Involving Decedents (CORID) at the University of Pittsburgh.

Specimens from four cerebral aneurysm domes were harvested during open brain surgery following surgical clipping in consented patients being treated for ruptured (n=1) and unruptured (n=3) aneurysms at the University of Illinois at Chicago (Chicago, IL) and Allegheny General Hospital (Pittsburgh, PA). Harvested specimens were placed in vials of HypoThermosol (Biolife, Bothell, USA), transported to the University of Pittsburgh in an insulated cooler where they were fixed within 24 hours of harvest in 4% paraformaldehyde. The intra-operative videos were also provided. The average patient age was 66±9 years old and the average specimen size was 5.6 ± 1.5 mm. The protocol for patient consent, handling of patient data, tissue harvest, and analysis were approved by the institutional review boards at the University of Illinois College of Medicine at Chicago, Allegheny General Hospital, and the University of Pittsburgh.

**Sectioned cadaveric specimens of cerebral artery: preparation and staining**

After harvest from the CoW, the MCA artery was immersed in optimal cutting temperature (OCT) compound and snap-frozen with liquid nitrogen until OCT compound completely solidified. The embedded sample was then sectioned with cryotome (Microm HM 525 Cryostat, Thermo Scientific, Germany) at 5 μm thickness. Three sectioned slices were placed on each slide with two slices used as negative controls for the primary and secondary antibodies, and one for full staining. Sectioned slices were twice washed for one minute with phosphate buffer solution (PBS). The frozen sections were first thawed at room temperature and then immersed in blocking reagent (5% normal goat serum (NGS)) for one hour at 37 ℃. Slides were then washed again with PBS, twice for 1 minute each. Monoclonal mouse anti-human smooth muscle actin clone 1A4 IgG2a antibody (Dako, Denmark) was used as the primary antibody and Alexa Fluor 488 goat anti-mouse IgG2a (Invitrogen, USA) for the secondary antibody. Primary antibody was diluted to a concentration of 1% with a 1% normal goat serum (NGS) staining buffer. Negative control and staining samples were stained with primary antibody solution for 45 minutes at 37 ℃. Stained slides were washed again with PBS, twice with 1 minute each. For secondary antibody, the solution was diluted to 1:100 ratio with 1% NGS. Negative controls for the secondary antibody and the staining sample were stained for 1 hour at 37℃. The slide was then washed with PBS, twice for 1 minute each and mounted with a mounting medium.

**Whole mount specimens: preparation and staining**

Cerebral artery and aneurysm specimens that were used for whole mount analysis were washed with PBS, twice for 1 minute each. The entire sample was then immersed into the same αSMA primary antibody used for sectioned tissue, for 6 hours at room temperature in the dark. The sample was then washed with PBS twice for 4 minutes each. Secondary antibody staining was performed for 3 hours at room temperature in the dark with the same secondary antibody as the sectioned specimens. Between staining, samples were thoroughly washed with PBS for 5 min.

**Imaging with confocal microscope**

Intact specimens of artery were scanned using a confocal microscope (Nikon A1, Tokyo Japan) with a lens magnification of 20x, an NA of 0.75 and a field of view (FoV) of 500 μm x 500 μm. The laser emission wavelengths were 450 nm for channel 1 and 525 nm for channel 2. Excitation frequencies were 405 nm for channel 1 and 488 nm for channel 2. Scanned images were reconstructed with IMARIS 9.5.0 (BitPlane AG, Zurich, Switzerland).

**Imaging with scanning multiphoton microscope**

A scanning multiphoton microscope (Nikon A1R MP HD, Tokyo, Japan) with the Nikon Ni-E upright motorized system and Chamelon Laser vision was used for large area scans of the artery and aneurysm specimens as well as for 500 μm x 500 μm scans of sectioned artery tissue (scanning feature not used). An APO LWD 25x water immersion objective lens with NA of 1.10 was used. The laser emission wavelength was set to 830 nm and the excitation frequencies were 400 to 492 nm for channel 1, and 500 to 550 nm for channel 2. For aneurysm and perforator artery scans, the large image scanning function was used with resonant scanning mode. A Nano-Drive system was used to acquire high-speed control of Z-plane selection. Scanned images were reconstructed with IMARIS 9.5.0 (BitPlane AG, Zurich, Switzerland). Virtual slices in the XZ plane and YZ plane (orthogonal to XY projected Z-stacks) were cut from the 3D reconstruction images with a thickness of 10 μm.

**Creation of in silico (virtual) model of harvested specimen and thickness maps**

Micro computed tomography (micro-CT) scans were conducted on fixed specimens using previously reported methods and the data sets were used to create in silico (virtual) 3D models of the samples and corresponding wall thickness maps.[12] Briefly, micro-CT scanning was performed using a high resolution scanner (Skycan 1272, Bruker Micro-CT, Kontich, Belgium) at a resolution of at least 3 μm. Virtual models of aneurysm samples were reconstructed from the Z stacks of micro-CT data (NRecon, Bruker Micro-CT, Kontich, Belgium). Wall thickness maps for each 3D virtual model were obtained using Materialise 3-matic (Materialise GmbH, Munich, Germany).

**Wall classification based on SMC actin density**

A cluster analysis was adapted to identify subregions within 2D projections of the 3D SMC actin MPM data sets based on SMC actin density. Here, subregions with high, medium and low density SMC actin were defined as Type 1, 2 and 3, respectively, Fig. 5. The raw SMC actin images, Fig. 5(a), were scaled down by a factor of 2, to speed up the evaluation. In Step 1, the 2D region occupied by the tissue specimen , Fig. 5(b) (R) within the entire projected SMC actin dataset was identified using thresholding (to remove the dark background), shown as the yellow bordered region in Fig. 5(c). In Step 2, subregions "devoid" of SMC actin were defined by color intensities lower than 20 percent of the brightest pixel visible within the entire tissue region, Fig. 5(d). The union of all such subregions was denoted by $R_0$. Type I subregions were then identified using the DBScan clustering algorithm,[39] which we previously implemented in [40]. In particular, in Step 3, high density regions within the sample were obtained by applying the clustering algorithm to the remaining region (R – $R_0$) using a density threshold of 12/7, Fig. 5(e). Here, the numerator corresponds to the number of pixels, and the denominator represents the neighborhood radius of the DBScan algorithm. The black cross markers in Fig.5 (f) show pixels that are identified as noise by the clustering algorithm. We then fit 2D surfaces to the selected pixels in this step by implementing the 2D triangulation algorithm, Fig.5(f). In Step 4, the Type 1 regions (collectively defined as region $R_1$, shown in green in Fig. 5(g)) were extracted by applying an area threshold of 400 (squared unit of pixels length scale) to these 2D regions, Fig. 5(h). The blue regions were excluded by surface area-based filtration, Fig 5(i). In Step 5, the same procedure described in Steps 3 and 4 was applied to the remaining region (R – $R_0$ – $R_1$) with a density threshold of 7/10 and area threshold of 2000 (squared unit of pixels length scale) to extract the Type 2 regions, collectively occupying region $R_2$. In Step 6, the Type 3 regions were then defined as the remaining regions ($R_3$ = R – $R_1$ – $R_2$), containing the low density SMC actin and regions identified as devoid of SMC actin ($R_0$). Color maps of the three wall types could then be created for each specimen, Fig. 5(j). The relative area of each wall type, could then be measured and used for sample quantification purposes.

**Mapping SI-MPM data to specimen model**

Prior to co-mapping the SMC actin and hemodynamic data, it ws necessary to map the SMC actin data to a 3D reconstructed virtual model of the harvested specimen (above), for registration purposes (ParaView 5.11.0, Kitware). First, the 2D SMC actin wall type data (png file) was resampled to reduce the resolution by 50% using the ResampleToImage filter in

ParaView. A planar surface mesh was created to obtain a 1-1 mapping between mesh location and wall type using the ExtractSurface filter. This mesh and associated data were then projected to the lumen side of the 3D virtual model of the specimen using the PointDatasetInterpolator filter. In Paraview, we used an Ellipsoidal Gaussian Kernel with a radius of 50, Sharpness of 20, with normal vectors option enabled. All three colors were then merged into a single "color" map on the lumen side of the model providing a map between SMC actin wall type and surface position on the lumen of the virtual model of the specimen.

**Computational fluid dynamic simulations (CFD) of blood flow and associated hemodynamic parameters**

Computational models for the 3D fluid domain were constructed from presurgical 3D rotational angiography (3DRA) images of the lumen of the aneurysm and neighboring vasculature using previously described techniques.[41] Blood was modeled as incompressible, constant viscosity (Newtonian) fluid with density $\rho$ = 1.0 g/cm$^3$ and viscosity $\mu$ = 0.04 Poise. Wall compliance was neglected, and no-slip boundary conditions were prescribed at the walls [42]. Pulsatile inflow boundary conditions were prescribed using flow waveforms measured in healthy subjects and scaled with a power law of the area of the inflow vessel. The split of the outflow was chosen, consistent with Murray's law were prescribed at the outlets. Periodic solutions to the coupled 3D system of nonlinear governing equations for unsteady flow of a Newtonian fluid (Navier-Stokes equations) and associated boundary conditions were solved numerically using finite elements with in-house software. Volumetric meshes composed of tetrahedral elements with a minimum resolution of 0.2 mm are generated filling the intravascular space using an advancing front method. Simulations were performed for 2 cardiac cycles with a timestep of 0.01 seconds, and results for pressure and velocity fields from the second cycle were saved for analysis. The velocity fields were used to calculate other previously defined field variables including the vortex core line and swirling flow around the vortex core lines.[43,44]

## Acknowldgements
We acknowledge the brain bank of the University of Pittsburgh for the cerebral arteries used in this work.
We acknowledge support for this work from the NINDS of the National Institutes of Health through grant 2R01NS097457.
We thank R. Fortunato (Department of Mechanical Engineering and Materials Science, University of Pittsburgh) and F. Mut for their advice on mapping the in silico models of the tissue specimens to the hemodynamic models.

## Author contributions
YT and AMR conceived of and designed the project. YT performed all imaging experiments and analysis of the associated data sets. SW provided expertise on the MPM studies. YT and AMR wrote the majority of the manuscript. MR developed and implemented the image segmentation algorithm for the clustering analysis of the SMC. MR wrote the corresponding parts of the methods section and associated figures. FC, SAH, AY performed surgeries and harvested aneurysm specimens. JRC designed and peformed the CFD studies and prepared the associated figures. All authors reviewed the manuscript.

## Competing interests
The authors declare no competing interests.